\documentclass[11pt,a4paper]{article}

\usepackage[paperwidth=26cm]{geometry}
\usepackage[T1]{fontenc}
\usepackage[latin1]{inputenc}
\usepackage{graphicx}
\usepackage{amsmath}
\usepackage{amssymb}

\newcommand{\bl}{\boldsymbol}

\newcommand{\lef}{\left(\,}
\newcommand{\rig}{\,\right)}
\newcommand{\ph}{\phantom}

\newcommand{\eq}{\,=\,}
\newcommand{\ma}{\,+\,}
\newcommand{\me}{\,-\,}

\begin{document}

\title{\textbf{Integrability Conditions for Killing-Yano Tensors and \\ Maximally Symmetric Spaces in the Presence of Torsion}}

\author{\textbf{Carlos Batista}\\
\small{Departamento de F\'{\i}sica}\\
\small{Universidade Federal de Pernambuco}\\
\small{50670-901 Recife-PE, Brazil}\\
\small{carlosbatistas@df.ufpe.br}}
\date{\today}




\maketitle

\begin{abstract}
The integrability conditions for the existence of Killing-Yano tensors or, equivalently, covariantly closed conformal Killing-Yano tensors, in the presence of torsion are worked out. As an application, all metrics and torsions compatible with the existence of a Killing-Yano tensor of order $n-1$ are obtained. Finally, the issue of defining a maximally symmetric space with respect to connections with torsion is addressed.
\textsl{(Keywords: Torsion, Integrability conditions, Conformal Killing-Yano tensors, Maximally symmetric spaces, General relativity)}
\end{abstract}

\section{Introduction}

In the standard model, the electromagnetic, weak, and strong interactions are all described by gauge theories for appropriate gauge groups. Therefore, it would be desirable to fit the fourth fundamental interaction, the gravitation, in this beautiful scheme. Although not obvious from the usual presentation, it turns out that general relativity can also be described as a gauge theory, with the Lorentz group being its structure group \cite{Utiyama}. Since the symmetry under spacetime translations is of fundamental physical relevance, it is natural to extend general relativity and look for the gauge theory whose structure is the Poincar\'{e} group. The  simplest gravitational theory with such a property is the so-called Einstein-Cartan theory \cite{Kibble&Sciama,Hehl-Rview}. The latter theory reduces to general relativity in vacuum. However, in the presence of particles with spin, Einstein-Cartan theory yields that the connection must be endowed with torsion,\footnote{If the matter field does not couple to the connection then no torsion is generated. For instance, the Lagrangian of the electromagnetic field is defined independently of the connection. So, in spite of the fact that such a field has spin one, it does not generate torsion.} departing from the formalism of general relativity. Since intrinsic spins are of quantum nature, it follows that gravitational theories with torsion may be of great relevance for the quantization of gravity. Particularly, in superstring theory, the field strength of the Kalb-Ramond field is generally interpreted as the torsion, as illustrated in \cite{Houri:Kerr-Sen}. In addition, torsion have also been considered in the context of AdS/CFT correspondence \cite{Leigh:AdS/CFT}, which can lead to applications in condensed matter physics.

One of the areas in which the physical implications of torsion have been exploited the most is cosmology.  For instance, the richness of the theories with torsion may be of relevance to explain dark matter and dark energy \cite{Puetzfeld-Review,Boehmer-CosmConst}. In addition, there are several works studying the possible connections between torsion and inflation \cite{Boehmer-Inflation,Puetzfeld-Review}. The torsion can also be used to change the apparent value of the cosmological constant, which can be valuable for a conciliation between supersymmetry and experimental observations  \cite{Boehmer-CosmConst} as well as for the solution of the so-called cosmological constant problem. A great drawback towards the acceptance of torsion as a useful tool to model our world is that it is very difficult to measure the effects of a torsion field. Indeed, it turns out that torsion couples just to particles with intrinsic spin \cite{Hehl71}. So, for instance, experiments like Gravity Probe B, in which the measuring devices are gyroscopes with macroscopic orbital angular momentum but with no net intrinsic spin, are not able to detect the possible existence of torsion \cite{Mao06,Puetzfeld-Probing}. In spite of such a difficulty, it has been pointed out that it is possible to put constraints in the torsion by means of the data from experiments of Lorentz violation \cite{Torsion-LorViol}. Actually, quite recently, an experiment using neutrons in liquid Helium have just been put forward with the aim of constraining the torsion field \cite{Lehnert14}.


The hidden symmetries represented by Killing-Yano (KY) tensors have proved to be of fundamental relevance to the development of four-dimensional general relativity. Indeed, it was due to the existence of a KY tensor of order two that the geodesic, Klein-Gordon, and Dirac equations could be analytically integrated in Kerr background \cite{Carter-constant,Walk-Pen,Carter-KleinG,Chandra-Dirac}. More generally, it has been proved that Kerr-NUT-(A)dS spacetimes of arbitrary dimension admit a closed conformal Killing-Yano tensor of rank two from which one can construct a \textit{tower} of KY tensors \cite{Frolov_KY} that provides just enough conserved charges to enable the explicit integration of the geodesic \cite{Kubiz-Geodesic}, the Klein-Gordon \cite{Frol-KG}, and the Dirac equations \cite{Oota-Dirac}. Furthermore, such hidden symmetries are also related to the separability of gravitational perturbations in Kerr-NUT-(A)dS spacetimes \cite{OotaPerturb}.

The present article aims to work out the integrability conditions for the existence of KY tensors, of arbitrary order, in the presence of torsion. To the best of author's knowledge, this has not been done before. These integrability conditions can be of great help in the involved task of integrating the KY equation, since they constrain the algebraic form of the KY tensor, which, in turn, eliminates several degrees of freedom in the general ansatz of a KY tensor, as illustrates \cite{Houri:2014} for the torsion-free case. The integrability conditions for KY tensors in the absence of torsion have already been obtained before in Refs. \cite{Kashiwada-Int,Bat-IntegraB-CKY}. Regarding works considering hidden symmetries in the presence of torsion, in Ref. \cite{Houri:Torsion} the metrics that allow the existence of a non-degenerate closed conformal Killing-Yano tensors of rank two in the presence of skew-symmetric torsion have been investigated, in Ref. \cite{Houri:TorsionDirac} it has been shown that, in general, a KY tensor does not lead to an operator that commutes with the Dirac operator if the connection has skew-symmetric torsion, whereas in Ref. \cite{Houri:Kerr-Sen} some spacetimes in which the torsion play an important role have been investigated. The interplay between KY tensors and torsion is also of relevance for the study of spaces with special holonomy \cite{Papadopoulos,Santillan}.

This article is organized as follows. Section \ref{Sec.Review} provides a review of connections with torsion. Particularly, the algebraic identities satisfied by the curvature tensor are displayed and a discussion about the ambiguity in the definition of geodesics and Killing vectors, due to the existence torsion,
is made. In Sec. \ref{Sec. KY and CCCKY}, we define KY tensors in the presence of torsion and show that they can be equivalently described as covariantly closed conformal Killing-Yano tensors. Then, the integrability conditions for the latter objects are obtained. As an application, in Sec. \ref{Sec.CCCKV}, all the metrics and antisymmetric torsions admitting a KY tensor of order $n-1$, with $n$ being the dimension of the space, are explicitly obtained. Finally, in Sec. \ref{Sec.MaximallySym}, the issue of defining a maximally symmetric space in the presence of torsion is investigated. In particular, it is shown that, for a general non-vanishing torsion, a manifold with the maximal number of Killing vectors does not admit the maximum number of KY tensors. For the sake of generality, most of the calculations performed here assume no constraint in the torsion such as skew-symmetry.

\section{Reviewing Connections with Torsion}\label{Sec.Review}

The intent of the present section is to provide a review of some basic aspects of the connections with torsion as well as to set the notation convention adopted in the forthcoming sections. A nice review on the role of torsion in theories of gravitation is available in \cite{Hehl-Rview}. Let $(M,\bl{g})$ be a differential manifold of dimension $n$ endowed with a metric tensor $\bl{g}$. Then, if $\bl{V}$ is a vector field in such manifold, its covariant derivative shall be written as
$$ \nabla_a\,V^b \eq \partial_a\,V^b \ma \Gamma^b_{ac}\,V^c \,,$$
where $\{a,b,c,\ldots\}$ denote coordinate indices and $\Gamma^b_{ac}$ is the connection symbol. Since the symbol $\Gamma^b_{ac}$ does not transform homogeneously under coordinate transformations, it turns out that it is not a tensor. Nevertheless, its skew-symmetric part in the lower indices is a tensor known as the Torsion tensor,
$$ T^{c}_{\ph{c}ab} \,\equiv\, 2\,\Gamma^c_{[ab]}  \eq\Gamma^c_{ab} \me \Gamma^c_{ba}  \,.$$
Where, as usual, indices enclosed by square brackets are anti-symmetrized whereas indices enclosed by round brackets are symmetrized. Because of its skew-symmetry in the last pair of indices it follows that the Torsion can be written as
\begin{equation}\label{Torsion-Decomposition}
  T_{abc}\eq \frac{2}{3}\,T_{(ab)c} \me \frac{2}{3}\,T_{(ac)b} \ma T_{[abc]}\,.
\end{equation}

Physically, it is reasonable to work with connections that preserve lengths and angles under parallel transports over the manifold, namely connections compatible with the metric. Therefore, we shall assume that the metric is covariantly constant, $\nabla_{a}g_{bc}\eq0$. One can solve the latter requirement for the connection and the solution is uniquely given by\footnote{A manifold endowed with this connection is called a Riemann-Cartan space.}
\begin{equation}\label{Connection-MetricCompatible}
 \Gamma^c_{ab} \eq \frac{1}{2}\,g^{ce}\,\left( \partial_{a}\,g_{be} \ma \partial_{b}\,g_{ae} \me \partial_{e}\,g_{ab} \me T_{abe} \me T_{bae}\ma T_{eab}\right) \eq \mathring{\Gamma}^c_{ab} \ma K^c_{\ph{c}ab}\,,
\end{equation}
with $\mathring{\Gamma}^c_{ab}$ denoting the familiar Levi-Civita connection, which is metric-compatible and torsion-free, whereas $K^c_{\ph{c}ab}$ is the so-called contortion tensor, whose definition is
\begin{equation}\label{ContortionDef}
  K^c_{\ph{c}ab} \,\equiv\,  \me \frac{1}{2}\,g^{ce}\,\left(  T_{abe} \ma T_{bae}\me T_{eab}\right)\,.
\end{equation}
It is interesting noting that, although the torsion arises from the anti-symmetric part of the connection, in general, it also contributes to the symmetric part of the connection. Indeed,
\begin{equation}\label{Connection-SymmetricPart}
\Gamma^c_{(ab)}\eq \mathring{\Gamma}^c_{ab} \me T_{(ab)}^{\ph{(ab)}c} \,.
\end{equation}

When we are dealing with a torsion-free connection, there are two equivalent ways of defining a geodesic: (i) Geodesics are the curves that minimize the distance between two points in the manifold; (ii) Geodesics are the integral curves of the vector fields with vanishing acceleration. However, for connections with non-zero torsion, these definitions cease to be equivalent. Indeed, if $\{x^a\}$ is a coordinate system, these two definitions lead to the following differential equations respectively:
\begin{align*}
  \delta\,s\,\eq \delta\int\sqrt{g_{ab}\,dx^a\,dx^b\,}\eq 0\, \quad\Longleftrightarrow\quad& \frac{d^2x^a}{ds^2}\ma \mathring{\Gamma}^a_{bc}\,\frac{dx^b}{ds}  \frac{dx^c}{ds} \eq 0 \\
  V^a\nabla_a\,V^b \eq 0 \quad\Longleftrightarrow\quad& \frac{d^2x^a}{ds^2}\ma \Gamma^a_{bc}\,\frac{dx^b}{ds}  \frac{dx^c}{ds} \eq 0\,,
\end{align*}
where $s$ represents the arc-length functional and $V^a=\frac{dx^a}{ds}$. Note that the first of the these differential equations can be written as $V^a\mathring{\nabla}_a\,V^b =0$, with $\mathring{\nabla}$ standing for the Levi-Civita covariant derivative. Due to Eq. (\ref{Connection-SymmetricPart}), it follows that these differential equations are equivalent if, and only if, $T_{(ab)c}=0$. Therefore, because of (\ref{Torsion-Decomposition}), we conclude these two definitions of a geodesic curve are equivalent if, and only if, the torsion is totally anti-symmetric, $T_{abc}=T_{[abc]}$. The case of totally skew-symmetric torsion is also of great relevance for string theory, in which the torsion is generally given by the field strength of a 2-form field. In spite of the physical appeal of the totally anti-symmetric torsions, in what follows most of the results are worked out without assuming this constraint.

It is worth remarking that none of these geodesics represent the orbits of general free falling particles in the presence of torsion. Indeed, if $V^a\mathring{\nabla}_aV^b=0$ had been the equation of motion of a general particle interacting just with the gravitational field then the particles would not have been affected by the torsion at all, so that it would have been physically unnecessary to introduce the concept of torsion. On the other hand, had $V^a\nabla_aV^b=0$ been the correct equation of motion of a general free falling particle then all particles would have been affected by the torsion. However, this cannot be true, since it is well established that only particles with intrinsic spin feel the torsion \cite{Yasskin80,Puetzfeld2008}. The requirement of diffeomorphism  invariance of the matter action implies a conservation law involving the energy momentum tensor. It turns out that by means of this conservation law it is possible to deduce the equations of motion of a test particle in a gravitational field, for an alternative derivation see \cite{Infeld}. For instance, in the torsion-less case the conservation law states that the divergence of the energy-momentum tensor vanishes and, using this fact, it is possible to obtain differential equations involving the momentum and the spin of the test particle \cite{Papapetrou,Dixon74}, the so-called Mathisson-Papapetrou-Dixon equations.\footnote{There are two inequivalent versions for these equations of motion, it turns out that the version deduced by Dixon implies the one obtained by Papapetrou but not the converse \cite{Khrapko}.} The analogous of these equations in the presence of torsion have been obtained in \cite{Hehl71,Yasskin80}. Nevertheless, it is worth pointing out that such equations of motion do not predict the whole motion of the test particle, since the dynamic of the center of mass of the particle is not fixed by these conservation laws. In spite of this, in the torsion-less case one can use the requirement of energy positiveness in order to argue that the momentum is proportional to the velocity of the center of mass \cite{Papapetrou,Khrapko} and then deduce that a point particle without intrinsic spin follow the geodesic path, but some ambiguities remain in the case of non-vanishing torsion \cite{Yasskin80}. However, it is worth stressing that, due to spin-orbit couplings, a point particle with non-zero spin generally will not follow the geodesic path even in the absence of torsion, an exception being the massless particles since in this case, semi-classically speaking, the spin must be aligned with the momentum \cite{PapapetrouII}. Irrespective of the correct path followed by a point test particle in the presence of torsion, the two concepts of geodesic presented here are of great geometrical significance and are worth studying in their own rights.

Whenever a fiber bundle is endowed with a connection, its curvature operator is defined by
$$ \nabla_{\bl{X}}\nabla_{\bl{Z}} - \nabla_{\bl{Z}}\nabla_{\bl{X}} - \nabla_{[\bl{X},\bl{Z}]} \,, $$
where $\bl{X}$ and $\bl{Z}$ are tangent vector fields and $[\bl{X},\bl{Z}]$ denotes their Lie Bracket. Particularly, assuming $\bl{X}$ and $\bl{Z}$ to be the coordinate vectors $\bl{\partial}_a$ and $\bl{\partial}_b$,  we end up with the following form for the curvature operator on the tangent bundle
$$ \left( \nabla_{a}\nabla_{b} \me \nabla_{b}\nabla_{a} \ma T^e_{\ph{e}ab}\nabla_{e}  \right) \,. $$
Since the action of the above operator in any scalar function gives zero, it follows that
$$ \left( \nabla_{a}\nabla_{b} \me \nabla_{b}\nabla_{a} \ma T^e_{\ph{e}ab}\nabla_{e}  \right)(f \bl{L}) \eq f \left( \nabla_{a}\nabla_{b} \me \nabla_{b}\nabla_{a} \ma T^e_{\ph{e}ab}\nabla_{e}  \right)\bl{L}  $$
for any tensorial field $\bl{L}$. Therefore, such operator defines a tensor called the curvature tensor and denoted by $R_{abc}^{\ph{abc}e}$. More precisely, the action of such operator in a tensor of rank $p$ is given by:
\begin{equation}\label{Riemann-Def}
  \left( \nabla_{a}\nabla_{b} \me \nabla_{b}\nabla_{a} \ma T^e_{\ph{e}ab}\nabla_{e}  \right)\,L_{c_1c_2\cdots c_p} \eq \sum_{i=1}^{p} R_{abc_i}^{\ph{abc_i}e}\,L_{c_1\cdots \check{c}_i \,e\, c_{i+1}\cdots c_p}\,,
\end{equation}
where $\check{c}_i$ means that the index $c_i$ has been withdrawn. Writing the covariant derivatives in the left hand side of (\ref{Riemann-Def}) in terms of partial derivatives and the connection symbol, we arrive at the following expression for the curvature:
\begin{equation}\label{Riemann-Gamma}
 R_{abc}^{\ph{abc}e} \eq \partial_{b}\Gamma^e_{ac} \me \partial_{a}\Gamma^e_{bc} \ma \Gamma^d_{ac} \Gamma^e_{bd} \me \Gamma^d_{bc} \Gamma^e_{ad} \eq  \me 2\, \partial_{[a}\Gamma^e_{b]c} \ma 2 \, \Gamma^d_{[a|c|} \Gamma^e_{b]d}\,.
\end{equation}
Now, inserting Eq. (\ref{Connection-MetricCompatible}) into the above relation, \textit{i.e.}, assuming the connection to be metric-compatible, one arrive at the following relation
\begin{equation}\label{Riemann-LeviC}
  R_{abc}^{\ph{abc}e} \eq \mathring{R}_{abc}^{\ph{abc}e} \me 2\, \nabla_{[a} K^e_{\ph{e}b]c} \ma 2\,  K^e_{\ph{e}[a|d|} K^d_{\ph{d}b]c} \me \,  T^d_{\ph{d}ab} K^e_{\ph{e}dc}\,,
\end{equation}
with $\mathring{R}_{abc}^{\ph{abc}e}$ standing for the curvature of the Levi-Civita connection. By means of the latter identity, one can prove that the
curvature of the general metric-compatible connection obeys the following identities:
\begin{align}
  R_{abcd} \eq& R_{[ab][cd]} \nonumber\\
 R_{[abc]}^{\ph{[abc]}e} \eq& -\,\nabla_{[a}T^e_{\ph{e}bc]} \ma T^d_{\ph{d}[ab}T^e_{\ph{e}c]d} \nonumber \\
 \nabla_{[a} R_{bc]de} \eq& T^f_{\ph{e}[ab}R_{c]fde}  \label{Riemann-Prop}\\
 R_{ab} \eq& R_{ba} \ma  \nabla_c T^c_{\ph{c}ab} \ma 2\,\nabla_{[a} T^c_{\ph{c}b]c} \ma T^c_{\ph{c}ab}T^d_{\ph{d}cd} \nonumber
\end{align}
Where $R_{ab}\equiv R^c_{\ph{c}acb}$ denotes the Ricci tensor. Note, in particular, that in the presence of torsion the Ricci tensor generally is not symmetric. In the special case in which the torsion is totally skew-symmetric, we further have the following relations:
\begin{equation}\label{Riemann-Prop2}
 T_{abc}=T_{[abc]}  \;\Longrightarrow\; \left\{
  \begin{array}{cl}
    R_{abcd} \eq& R_{cdab} \ma \nabla_{[c}T_{d]ab} \me \nabla_{[a}T_{b]cd}   \\
    R_{ab} \eq& R_{ba} \ma \nabla_c T^c_{\ph{c}ab}\,. \\
    \end{array}
\right.
\end{equation}

%
%

The exterior derivative of a differential form can be defined irrespective of the existence of a connection on the bundle of differential forms. Indeed, if $\bl{F}$ is a $p$-form then its exterior derivative is a $(p+1)$-form whose components are:
$$  (dF)_{ab_1b_2\cdots b_p} \eq (p+1)\,\partial_{[a}F_{b_1b_2\cdots b_p]} \,.$$
It is well-known that in the case of a torsion-free connection it is harmless to replace the partial derivative in the latter expression by a covariant derivative. Nevertheless, for connections with torsion this replacement is not allowed anymore. Instead, the following relation holds:
$$ (dF)_{ab_1b_2\cdots b_p} \eq (p+1)\,\nabla_{[a}F_{b_1b_2\cdots b_p]} \ma \frac{p(p+1)}{2}\,F_{c[b_2\cdots b_p}\,T^c_{\ph{c}ab_1]} \,. $$
As usual, if $d\bl{F}=0$ we shall say that the differential form $\bl{F}$ is closed, whereas if $\nabla_{[a}F_{b_1b_2\cdots b_p]}=0$ then $\bl{F}$ will be said to be covariantly closed. These concepts coincide just for connections with vanishing torsion.

\subsection{Killing Vectors and Torsion}\label{SubSec-KillingV}

In the absence of torsion there are two equivalent ways of saying that a vector field $\bl{\eta}$ is a Killing vector: (i) The metric is invariant under the Lie dragging along the orbits of $\bl{\eta}$, namely $\mathcal{L}_{\bl{\eta}}g_{ab} =0$, with $\mathcal{L}_{\bl{\eta}}$ standing for the Lie derivative along $\bl{\eta}$; (ii) The vector field $\bl{\eta}$ obeys the Killing equation $\nabla_{(a}\eta_{b)}=0$. However, if the connection has torsion these two definitions generally are not equivalent anymore. Indeed, one can check that the following relation holds:
\begin{equation}\label{Killings-Torsion}
  \nabla_{a}\eta_{b}\ma  \nabla_{b}\eta_{a} \eq \mathcal{L}_{\bl{\eta}}g_{ab} \ma 2 \,T_{(ab)e}\,\eta^e \,.
\end{equation}
If we choose a coordinate frame in which the Killing vector is one of the basis vectors, $\bl{\eta}=\bl{\partial}_{x^1}$, then, in these coordinates, the operator $\mathcal{L}_{\bl{\eta}}$ is given by the partial derivative $\partial_{x^1}$. Therefore, the definition (i) means that the components of the metric do not depend on the coordinate $x^1$. Differently, in the presence of torsion, the existence of a vector field obeying $\nabla_{(a}\eta_{b)}=0$ does not guarantee that the metric is independent some coordinate in a suitable coordinate frame.

These two notions of Killing vector are intimately related to the two distinct definitions of geodesic. In fact, if $\bl{V}$ is a vector field tangent to a geodesic and $\bl{\eta}$ is a Killing vector according to the definition (i) then the scalar $(\eta_aV^a)$ is conserved along the curves of minimum length, while if $\bl{\eta}$ is a Killing vector according to the definition (ii) then $(\eta_a V^a)$ is conserved along the curves of zero acceleration. More precisely,
\begin{align*}
  V^a\mathring{\nabla}_a\,V^b =0 \quad\textrm{ and }\quad \mathcal{L}_{\bl{\eta}}g_{ab} =0 \quad&\Longrightarrow\quad V^a\nabla_a(\eta_b V^b) \eq 0 \,,\\
 V^a\nabla_a\,V^b =0 \quad\textrm{ and }\quad \nabla_{(a}\eta_{b)}=0 \quad&\Longrightarrow\quad V^a\nabla_a(\eta_b V^b) \eq 0\,.
\end{align*}
Looking at Eq. (\ref{Killings-Torsion}), one conclude that in order for both definitions of a Killing vector to be equivalent we must have $T_{(ab)c}=0$, which means that the torsion is totally skew-symmetric. Likewise, this is exactly the necessary condition for the two concepts of geodesic to be equivalent. In what follows we shall mainly stick to the definition (ii) of a Killing vector.

Before proceeding to the integrability conditions of the conformal Killing-Yano tensors, let us write the second derivative of a Killing vector in a convenient way for later purposes. Here, by a Killing vector it is meant a vector field obeying the Killing equation with the general metric-compatible connection. Thus, if $\bl{\eta}$ is a Killing vector then the following equation hold:
$$ \nabla_{a}\nabla_{b}\eta_c \ma \nabla_a \nabla_c \eta_b\eq \nabla_{a}\lef \nabla_{b}\eta_c \ma  \nabla_c \eta_b \rig \eq 0 \,. $$
Using permutations of this equation along with (\ref{Riemann-Def}) we find
\begin{align}
  2\,\nabla_{a}\nabla_{b}\eta_c &\eq (\nabla_{a}\nabla_{b}-\nabla_{b}\nabla_{a})\eta_c \ma (\nabla_{c}\nabla_{a}-\nabla_{a}\nabla_{c})\eta_b \ma (\nabla_{c}\nabla_{b}-\nabla_{b}\nabla_{c})\eta_a  \nonumber \\
  &\eq (R_{abc}^{\ph{abc}e}+R_{cab}^{\ph{cab}e}+R_{cba}^{\ph{abc}e})\eta_e\me (T^e_{\ph{e}ab}\nabla_e\eta_c+ T^e_{\ph{e}ca}\nabla_e\eta_b+ T^e_{\ph{e}cb}\nabla_e\eta_a)  \nonumber \\
  &\eq \left( 3\,R_{[abc]}^{\ph{[abc]}e}\ma 2\,R_{cba}^{\ph{cba}e} \right)\eta_e\me (T^e_{\ph{e}ab}\nabla_e\eta_c+ T^e_{\ph{e}ca}\nabla_e\eta_b+ T^e_{\ph{e}cb}\nabla_e\eta_a) \,.\label{KillingV-SecondD}
\end{align}

\section{Killing-Yano and Covariantly Closed Conformal Killing-Yano Tensors}\label{Sec. KY and CCCKY}

In this section we shall define Killing-Yano tensors and covariantly closed conformal Killing-Yano tensors and argue that these objects are two sides of the same coin. Then, the integrability conditions for the existence of these objects will be obtained.

A non-zero $p$-form $\bl{Y}$ is called a conformal Killing-Yano (CKY) tensor of order $p$ whenever it obeys the following differential equation
\begin{equation}\label{CKY1}
  \nabla_{a}\, Y_{b_1b_2\cdots b_p} \ma \nabla_{b_1}\, Y_{ab_2\cdots b_p} \eq 2\,g_{a[b_1}\,h_{b_2\cdots b_p]} \ma 2\,g_{b_1[a}\,h_{b_2\cdots b_p]} \,,
\end{equation}
with $h_{b_2\cdots b_p}$ being some totaly anti-symmetric tensor of rank $p-1$. Actually, contracting the above equation with $g^{ab_1}$ we find that
\begin{equation}\label{h=Divergence}
   h_{b_2\cdots b_p}  \eq \frac{p}{2(n+1-p)}\,\nabla^a\,Y_{ab_2\cdots b_p} \,,
\end{equation}
with $n$ standing for the dimension of the manifold. By means of algebraic manipulations, one can show that the conformal Killing-Yano equation (\ref{CKY1}) is equivalent to the following condition:
\begin{equation}\label{CKY2}
  \nabla_{a}\, Y_{b_1b_2\cdots b_p} \eq  \nabla_{[a}\, Y_{b_1b_2\cdots b_p]} \ma 2\,g_{a[b_1}\,h_{b_2\cdots b_p]} \,.
\end{equation}
There are two very special cases of CKY tensors. If $h_{b_2\cdots b_p}$ vanishes, \textit{i.e.}, if $\bl{Y}$ has vanishing divergence, then the CKY tensor is called a Killing-Yano (KY) tensor. While if $\nabla_{[a}\, Y_{b_1b_2\cdots b_p]}$ vanishes then we shall say that $\bl{Y}$ is a covariantly closed conformal Killing-Yano (CCCKY) tensor.

It turns out that every KY tensor is the Hodge dual of a CCCKY tensor and \textit{vice versa}. Indeed, if $\epsilon_{a_1a_2\cdots a_n}$ is the volume-form of the manifold then it follows that
\begin{equation}\label{EE_DELTA}
  \epsilon^{a_1\ldots a_q\,b_{q+1}\ldots b_n}\,\epsilon_{a_1\ldots a_q\,c_{q+1}\ldots c_n}\;=\;q!(n-q)!\, (-1)^{\frac{n-s}{2}} \delta_{c_{q+1}}^{\;[b_{q+1}}\ldots\delta_{c_{n}}^{\;b_{n}]}\,,
\end{equation}
with $s$ being the signature of the metric. Taking the covariant derivative of the latter equation we conclude that $\bl{\epsilon}$ is covariantly constant with respect to any metric-compatible connection. Now, let $\bl{A}$ be a KY tensor of rank $n-p$, so that
\begin{equation}\label{KY-Def}
  \nabla_a A_{b_1\cdots b_{n-p}}  \eq  \nabla_{[a}\, A_{b_1b_2\cdots b_{n-p}]}\,.
\end{equation}
Then, let us define the tensors
$$  H^{a_1a_2\cdots a_p}  \,\equiv\,   \epsilon^{a_1a_2\cdots a_p b_1\cdots b_{n-p}}\,A_{b_1\cdots b_{n-p}} \quad  \textrm{and}     $$
$$
    h^{a_1a_2\cdots a_{p-1}}  \,\equiv\,  \frac{ (-1)^{(p-1)}\,p!\, (n-p)!}{(p-1)!\, (n-p+1)!} \, \epsilon^{a_1a_2\cdots a_{p-1} c b_1\cdots b_{n-p}} \,\nabla_{[c}\, A_{b_1b_2\cdots b_{n-p}]} \,. $$
Now, contracting the above definition of $h^{a_1a_2\cdots a_{p-1}}$ with $\epsilon_{a_1a_2\cdots a_{p-1}d e_1\cdots e_{n-p}}$ we find that
$$  \nabla_{[a}\, A_{b_1b_2\cdots b_{n-p}]} \eq \frac{ (-1)^{(\frac{n-s}{2}+p-1)} }{p!\, (n-p)!} \, h^{c_1c_2\cdots c_{p-1}}\epsilon_{c_1c_2\cdots c_{p-1}ab_1b_2\cdots b_{n-p}}\,. $$
Inserting the latter identity into (\ref{KY-Def}) then contracting the final equation with $\epsilon^{d_1d_2\cdots d_p b_1\cdots b_{n-p}}$, and using the fact that the volume form is covariantly constant, eventually lead us to the following differential equation:
$$   \nabla_a H_{b_1\cdots b_p} \eq 2\, g_{a[b_1} h_{b_2\cdots b_p]} \,. $$
Thus, $\bl{H}$ is a CCCKY tensor. Since, apart from a non-important multiplicative constant,  $\bl{H}$ is the Hodge dual of $\bl{A}$, we have proved that the Hodge dual of a KY tensor is a CCCKY tensor. In a completely analogous fashion, one can prove that the Hodge dual of every CCCKY tensor is a KY tensor. Hence, studying the CCCKY equation is equivalent to analysing KY equation. Particularly, in this article we have made the choice of working out the integrability conditions for CCCKY tensors of arbitrary rank. The choice of dealing with CCCKY tensors instead of KY tensors is mainly based on the fact that in Kerr-NUT-(A)dS spacetimes of arbitrary dimension a CCCKY tensor of rank two is the origin of the integrability of Klein-Gordon and Dirac equations in this background \cite{Frol-KG,Oota-Dirac}.

The importance of Killing-Yano tensors relies on the fact that they generate conserved quantities along the geodesic curves, where here a geodesic means a curve of zero acceleration. Indeed, if $\bl{V}$ is a geodesic vector field, $V^a\nabla_aV^b=0$, and $\bl{A}$ is a KY tensor of order $p$ then it follows that the scalar
$$ C \eq V^a\,V^b\, A_{ac_2\cdots c_p}\,A_{b}^{\ph{b}c_2\cdots c_p}  $$
is conserved along the geodesic tangent to $\bl{V}$, \textit{i.e.}, $V^a\nabla_aC=0$. In particular, this implies that the symmetric tensor
$$ Q_{ab}\eq  A_{ac_2\cdots c_p}\,A_{b}^{\ph{b}c_2\cdots c_p}   $$
is a Killing tensor of rank two, namely $\nabla_{(a}Q_{bc)}\eq 0$. More generally, if $\bl{A}$ and $\widehat{\bl{A}}$ are both KY tensors of order $p$ then
$$ \widetilde{Q}_{ab}\eq  A_{(a}^{\ph{a}c_2\cdots c_p} \,\widehat{A}_{b)c_2\cdots c_p}  $$
is a Killing tensor of rank two. On the other hand, the utility of CCCKY tensors relies on the fact that the exterior product of two CCCKY tensors is another CCCKY tensor. Thus, if $\bl{H}$ and $\widehat{\bl{H}}$ are CCCKY tensors of order $p$ and $q$ respectively then
$$ \widetilde{H}_{a_1\cdots a_p b_1\cdots b_q} \eq H_{[a_1\cdots a_p }\,\widehat{H}_{b_1\cdots b_q] }  $$
is a CCCKY tensor of order $(p+q)$. Therefore, once we a have a CCCKY tensor we can, in principle, build a tower of CCCKY tensors by means of taking exterior products of the CCCKY tensor with itself. Then, taking the Hodge dual of these CCCKY tensors we end up with a tower of KY tensors and, hence, a tower of conserved scalars. Note that, since Killing tensors lead to conserved quantities along geodesics, it follows from Noether's theorem that such tensors might generate symmetry transformations that leave particle's action invariant, a fact that have been addressed in Refs. \cite{Santillan,Holten-Symmetries}. In the case of KY tensors they also generate symmetries in superspace \cite{Santillan}. For more on the relation between KY tensors and conserved quantities as well as their importance in general relativity, the reader is referred to Ref. \cite{Frolov_KY} and references therein.

\subsection{Integrability Condition}\label{SubSec.IntegCond}

Now let us obtain the integrability conditions for the existence of a covariantly closed conformal Killing-Yano tensor of arbitrary order in the presence of torsion. The torsion-free case have already been addressed in \cite{Kashiwada-Int,Bat-IntegraB-CKY}. A skew-symmetric tensor $\bl{H}$ of rank $p$ is said to be a CCCKY tensor if it obeys the following equation:
\begin{equation}\label{CCCKY-Def}
  \nabla_a H_{b_1\cdots b_p} \eq 2\, g_{a[b_1} h_{b_2\cdots b_p]} \,,
\end{equation}
where $\bl{h}$ is a skew-symmetric tensor of rank $(p-1)$ given by
\begin{equation}\label{h=Divergence2}
   h_{b_2\cdots b_p}  \eq \frac{p}{2(n+1-p)}\,\nabla^a\,H_{ab_2\cdots b_p} \,.
\end{equation}
In order to simplify the notation, let us define the tensor $\bl{h}'$ as
$$ h'_{ab_2\cdots b_p} \eq \nabla_a h_{b_2\cdots b_p}\,.$$
Particularly, note that $h'_{ab_2\cdots b_p}=h'_{a[b_2\cdots b_p]}$. Now, let us use Eq. (\ref{Riemann-Def}) in order to compute the trace of $\bl{h}'$:
\begin{align}\label{h'aa}
  h'^{a}_{\ph{a}ab_3\cdots b_p} \eq& \nabla^a \, h_{ab_3\cdots b_p} \eq \frac{p}{2(n+1-p)}\,\nabla^a\nabla^c\,H_{cab_3\cdots b_p} \nonumber \\
 \eq& \frac{-\,p}{4(n+1-p)}\,(\nabla^a\nabla^c-\nabla^c\nabla^a)\,H_{acb_3\cdots b_p} \nonumber\\
 \eq& \frac{p}{4(n+1-p)}\left[ 2\,R^{ac}\,H_{acb_3\cdots b_p} + \sum_{i=3}^p R^{ace}_{\ph{[ace]}b_i}\,H_{acb_3\cdots \check{b}_i \,e\, b_{i+1}\cdots b_p}\ma T^{eac}\nabla_e H_{acb_3\cdots b_p} \right]\nonumber\\
 \eq& \frac{p}{4(n+1-p)}\left[ 2\,R^{[ac]}\,H_{acb_3\cdots b_p} + \sum_{i=3}^p R^{[ace]}_{\ph{[ace]}b_i}\,H_{acb_3\cdots \check{b}_i \,e \,b_{i+1}\cdots b_p}\right. \nonumber\\
   & \ph{\frac{p}{4(n+1-p)}}\;\;\,\ma\, \left. \frac{4}{p}\,T_a^{\ph{a}ac}h_{cb_3\cdots b_p}\ma \frac{2(p-2)}{p}\,h_{ac[b_4\cdots b_p}T_{b_3]}^{\ph{b_3]}ac} \right]
\end{align}
Where it is worth recalling that the notation $\check{b}_i$ means that the index $b_i$ is absent. In order to attain (\ref{h'aa}) it has been used the following useful algebraic identity valid for any totally skew-symmetric tensor $\bl{A}$ of rank $(p-1)$:
\begin{align}\label{AlegraicIdentity}
  g_{e[a}A_{cb_3\cdots b_p]}\, \eq& \frac{1}{p}\, g_{ea} \,A_{cb_3\cdots b_p} \me \frac{p-1}{p}\,  \,A_{a[b_3\cdots b_p}g_{c]e}  \nonumber \\
 \eq& \frac{1}{p}\, g_{ea} \,A_{cb_3\cdots b_p} \me \frac{1}{p}\, g_{ec} \,A_{ab_3\cdots b_p} \ma \frac{p-2}{p}\,  \,A_{ac[b_4\cdots b_p}g_{b_3]e}\,.
\end{align}
Note that the right hand side of (\ref{h'aa}) is zero for vanishing torsion. Now, taking the covariant derivative of (\ref{CCCKY-Def}), we find that
$$  \nabla_a\nabla_b H_{c_1c_2\cdots c_p} \eq 2\, h'_{a[c_2\cdots c_p}\,g_{c_1]b} $$
Thus, using this equation along with (\ref{Riemann-Def}) we arrive at
\begin{align}\label{Int1}
 2\, h'_{a[c_2\cdots c_p}\,g_{c_1]b} \me 2\, h'_{b[c_2\cdots c_p}\,g_{c_1]a}  \eq& ( \nabla_a\nabla_b\me  \nabla_b\nabla_a) H_{c_1c_2\cdots c_p} \nonumber\\
  \eq& \sum_{i=1}^p R_{abc_i}^{\ph{abc_i}e}\,H_{c_1c_2\cdots \check{c}_i \,e\, c_{i+1}\cdots c_p} \me T^e_{\ph{e}ab}\nabla_e H_{c_1c_2\cdots c_p} \nonumber\\
  \eq& \sum_{i=1}^p R_{abc_i}^{\ph{abc_i}e}\,H_{c_1c_2\cdots \check{c}_i \,e\, c_{i+1}\cdots c_p} \me 2\,h_{[c_2\cdots c_p}T_{c_1]ab} \,.
\end{align}
As it is, this equation is an integrability condition involving the CCCKY tensor $\bl{H}$ and its second derivative through the tensor $\bl{h}'$. However, in order to apply an integrability condition it is more useful when it is purely algebraic. Therefore, let us try to express the tensor $\bl{h}'$ in terms of $\bl{H}$ and $\bl{h}$. Expanding Eq. (\ref{Int1}) by means the algebraic identity (\ref{AlegraicIdentity}) lead us to
\begin{align*}
 \frac{2}{p}\,g_{c_1b} & \,h'_{ac_2\cdots c_p}\me \frac{2(p-1)}{p}\ h'_{ac_1[c_3\cdots c_p}g_{c_2]b} \me  \frac{2}{p}\,g_{c_1a}\, h'_{bc_2\cdots c_p}\ma \frac{2(p-1)}{p}\ h'_{bc_1[c_3\cdots c_p}g_{c_2]a}  \nonumber\\
  \eq&  R_{abc_1}^{\ph{abc_1}e}\,H_{ec_2\cdots  c_p}\ma \sum_{i=2}^p R_{abc_i}^{\ph{abc_i}e}\,H_{c_1c_2\cdots \check{c}_i \,e\, c_{i+1}\cdots c_p} \me  \frac{2}{p}\,T_{c_1ab}\,h_{c_2\cdots c_p} \ma \frac{2(p-1)}{p}h_{c_1[c_3\cdots c_p}T_{c_2]ab} \,.
\end{align*}
So, contracting the latter equation with $g^{bc_1}$, we end up with the following expression for $\bl{h}'$:
\begin{align}\label{h'Algeb}
  h'_{ac_2\cdots c_p} \eq & \frac{-\,p}{2(n-p)}\left[ R_a^{\ph{a}e}\,H_{ec_2\cdots c_p} \ma \sum_{i=2}^p R_{\ph{b}ac_i}^{b\ph{ac_i}e}\,
  H_{bc_2\cdots \check{c}_i e\cdots c_p} \ma \frac{2(p-1)}{p}\, h'^{e}_{\ph{e}e[c_3\cdots c_p}g_{c_2]a}\right.  \nonumber \\
 &  \ph{\frac{-\,p}{2(n-p)}} \quad \ma \left.\frac{2(p-1)}{p}\, h^e_{\ph{e}[c_3\cdots c_p}T_{c_2]ea} \me \frac{2}{p} \,h_{c_2\cdots c_p}\, T^e_{\ph{e}ea} \right]
\end{align}
Then, inserting (\ref{h'aa}) into (\ref{h'Algeb}) we can find the wanted expression for $\bl{h}'$. Finally, inserting such expression into (\ref{Int1}) we arrive at the integrability condition for the existence of a CCCKY tensor involving just $\bl{H}$ and $\bl{h}$, without higher derivatives of $\bl{H}$. Thus, the integrability condition amounts to the Eqs. (\ref{Int1}), (\ref{h'Algeb}) and (\ref{h'aa}).

As a first consequence of such integrability condition, let us obtain a constraint on the torsion tensor. Taking the totally skew-symmetric part of indices $abc_1\cdots c_p$ in Eq. (\ref{Int1}) we find that
\begin{equation}\label{TorsionInt1}
   2\,T_{[abc_1} \,h_{c_2\cdots c_p]} \eq \sum_{i=1}^p R_{[abc_i}^{\ph{[abc_i}e}\,H_{c_1c_2\cdots \check{c}_i |e|c_{i+1}\cdots c_p]} \eq (-1)^{(p-1)}\,p\,R_{[abc_1}^{\ph{[abc_1}e}\,H_{c_2\cdots c_p]e} \,.
\end{equation}
Now, using the first Bianchi identity in (\ref{Riemann-Prop}) we can write the above equations as
\begin{equation}\label{TorsionInt2}
   2\,T_{[abc_1} \,h_{c_2\cdots c_p]} \eq (-1)^{p}\,p\,\left(\, \nabla_{[a}T^e_{\ph{e}bc_1} \,H_{c_2\cdots c_p]e}
   \ma    T^e_{\ph{e}d[a}T^d_{\ph{d}bc_1}H_{c_2\cdots c_p]e} \,\right) \,.
\end{equation}
This is a constraint involving just the torsion. Therefore, besides the metric, the torsion is also constrained by the existence of a covariantly closed conformal Killing-Yano tensor.

As another consequence of these integrability conditions, let us consider the case $p=2$, namely a bivector $\bl{H}$ obeying to $\nabla_a H_{bc} = 2 g_{a[b} h_{c]}$. It is well known that in the absence of torsion the vector field $h_a$ is a Killing vector whenever the Ricci tensor is proportional to the metric \cite{Bat-IntegraB-CKY}. Differently, since $\nabla_{(a}h_{b)}=h'_{(ab)}$ it follows from (\ref{h'Algeb}) that if torsion is different from zero then, in general, $h_a$ will not be a Killing vector even if the Ricci tensor is proportional to the metric. For instance, if the torsion is totally skew-symmetric and $R_{ab}\propto g_{ab}$ then Eqs. (\ref{h'Algeb}), (\ref{h'aa}), (\ref{Riemann-Prop}) and (\ref{Riemann-Prop2}) imply that $\nabla_{(a}h_{b)}$ vanishes if, and only if,
$$  \lef  \nabla_{[a}T_{e]cb} + \nabla_{[c}T_{e]ab} \rig\,H^{eb} $$
also vanishes, which generally is not the case. Now, let us apply the results obtained in the present section for the case $p=1$, \textit{i.e.}, the case of a covariantly closed conformal Killing vector.

\section{Covariantly Closed Conformal Killing Vectors}\label{Sec.CCCKV}

The aim of the present section is to find all metrics and torsions compatible with the existence of a covariantly closed conformal Killing vector (CCCKV), namely a vector field $\bl{H}$ obeying
\begin{equation}\label{CCCKV}
  \nabla_a\,H_b \eq 2\,h\, g_{ab} \quad \textrm{ with } \quad h \eq \frac{1}{2\,n}\,\nabla_aH^a \,.
\end{equation}
Since every Killing-Yano tensor of order $n-1$ is the Hodge dual of a CCCKV, we will, equivalently, find all metrics and torsions compatible with a KY tensor of order $n-1$.

Assuming $p=1$ in Eqs. (\ref{Int1}) and (\ref{TorsionInt2}) we arrive at the following integrability conditions:
\begin{equation}\label{Int-CCCKV1}
  R_{abc}^{\ph{abc}e}H_e \ma 2\,g_{ca}\,h'_b \me 2\,g_{cb}\,h'_a \eq 2\,h\,T_{cab} \,,
\end{equation}
\begin{equation}\label{Int-CCCKV2}
  H_{e}\,\nabla_{[a}T^e_{\ph{e}bc]}\ma  H_{e}\, T^e_{\ph{e}d[a}T^d_{\ph{d}bc]} \ma  2\,h\,T_{[abc]} \eq 0 \,.
\end{equation}
Where, using (\ref{h'Algeb}), we have that $h'_a$ is given by
\begin{equation}\label{h'-CCCKV}
 h'_a\eq \nabla_a h \eq \frac{-\,1}{2(n-1)}\, \left( \, R_a^{\ph{a}e}\,H_e \me 2\,h\,T^e_{\ph{e}ea}   \, \right) \,.
\end{equation}
Therefore, inserting (\ref{h'-CCCKV}) into (\ref{Int-CCCKV1}) we arrive at the following algebraic integrability condition:
\begin{equation}\label{Int-CCCKV-Algeb}
  \left[\,(n-1)R_{abc}^{\ph{abc}e} \me g_{ac}\,R_b^{\ph{b}e} \ma g_{bc}\,R_a^{\ph{a}e} \,\right]\,H_e \eq
  2\,h\,\left[\,(n-1)\,T_{cab}\me g_{ac}\,T^e_{\ph{e}eb} \ma  g_{bc}\,T^e_{\ph{e}ea}  \,\right] \,.
\end{equation}

Now, let us denote the squared norm of $\bl{H}$ by $N$,
$$ N \,\equiv\, H_a\,H^a \,.  $$
Then, taking the derivatives of this equation and using (\ref{CCCKV}) we find that
\begin{align}
   \nabla_b N \eq&   4\,h\,H_b \,,\label{DN} \\
  \nabla_a \nabla_b N \eq& 4\,h'_a\,H_b \ma 8\,h^2\,g_{ab}\,. \label{DDN}
\end{align}
The commutator of covariant derivatives acting in a scalar gives
$$ (\nabla_a \nabla_b \me \nabla_b \nabla_a)N \eq -\,T^e_{\ph{e}ab}\nabla_e N \,. $$
Thus, inserting (\ref{DN}) and (\ref{DDN}) into this identity we find the following useful relation
\begin{equation}\label{Hh'}
  H_a\, h'_b \me H_b\, h'_a \eq h\,H_e\, T^e_{\ph{e}ab} \,.
\end{equation}
In particular, such equation implies
\begin{equation}\label{HyperSurfOrt1}
  H_e\, T^e_{\ph{e}[ab}\,H_{c]}\eq  0 \,\,,  \quad\; \textrm{ if $h\neq 0$}\,.
\end{equation}
Note that $h\neq0$ means that the vector $\bl{H}$ is not covariantly constant. Now, since $\bl{H}$ is covariantly closed it follows that
\begin{equation}\label{APDH}
   0\eq \nabla_{[a}H_{b]} \eq \partial_{[a}H_{b]} - \Gamma^e_{[ab]}H_e \quad \Longrightarrow \quad \quad
    \partial_{[a}H_{b]} \eq \frac{1}{2}\,H_e\,T^e_{\ph{e}ab} \,.
\end{equation}
So, using (\ref{APDH}) and (\ref{HyperSurfOrt1}) we conclude that
\begin{equation}\label{HyperSurfOrt2}
 H_{[c}\,\partial_{a}H_{b]} \eq 0 \,\,,  \quad\; \textrm{ if $h\neq 0$}\,.
\end{equation}

In what follows we shall assume the condition $h\neq 0$ to hold, the case $h=0$ will be considered separately later. According to the Frobenius theorem, the relation (\ref{HyperSurfOrt2}) guarantees that the vector field $\bl{H}$ is orthogonal to a family of hyper-surfaces. Thus, if $\lambda$ is a parameter along the orbits of $\bl{H}$ we can locally introduce coordinates $\{x^i,\lambda\}$, with $i,j \in\{1,2,\cdots,n-1\}$, such that the vector field
$\bl{H}= \bl{\partial}_\lambda$ is orthogonal to the $(n-1)$ basis vectors $\bl{\partial}_i$. Therefore, in these coordinates the line element is given by
\begin{equation}\label{Metric1}
  ds^2 \eq N\,d\lambda^2 \ma g_{ij}\,dx^i dx^j \,.
\end{equation}
Where, in principle, $N$ and $g_{ij}$ are functions of $\lambda$ and $\{x^i\}$. However, since in these coordinates the components of $\bl{H}$ are
\begin{equation*}\label{H-Components}
  H^a\eq \delta^a_\lambda  \quad\textrm{ and }\quad H_{a}\eq N\,\delta^\lambda_a
\end{equation*}
it follows from (\ref{DN}) that
\begin{equation}\label{PDN}
  \partial_aN = 4\,h\,N\,\delta^\lambda_a \quad \Longrightarrow \quad \left\{
  \begin{array}{ll}
     \partial_\lambda N \eq 4\,h\,N\,,\\
    \partial_iN \eq 0 \;\; \Rightarrow \;\; N = N(\lambda)\,.\\
    \end{array}
\right.
\end{equation}
Therefore, $N$ is just a function of $\lambda$. Consequently,  we have that
\begin{equation}\label{H-closed}
  \partial_{[a}H_{b]} \eq  \partial_{[a}N \delta^\lambda_{b]}  \eq \partial_\lambda N \, \delta^\lambda_{[a} \delta^\lambda_{b]} \eq 0\,.
\end{equation}
Thus, besides being covariantly closed, it turns out that $\bl{H}$ is also closed. As a consequence of (\ref{H-closed}) and (\ref{APDH}), we arrive at the following constraint for the torsion:
\begin{equation}\label{Tlambda ab}
  H^e\,T_{eab} \eq T_{\lambda ab} \eq  0 \,.
\end{equation}
This, along with (\ref{Hh'}), implies that
\begin{equation}\label{h'-prop-H}
  h'_a \,\propto\, H_a    \quad \Longrightarrow \quad  h=h(\lambda)\,,
\end{equation}
which could be anticipated from (\ref{PDN}). In addition, taking the covariant derivative of (\ref{Tlambda ab}) and then using (\ref{CCCKV}) we immediately find
\begin{equation}\label{DT-H}
  H_e \nabla_c T^e_{\ph{e}ab} \eq -\, 2\,h\, T_{cab} \,.
\end{equation}
The latter identity along with (\ref{Tlambda ab}) implies that the torsion integrability condition (\ref{Int-CCCKV2}) is readily satisfied. Now, writing down the equation $\nabla_aH_b=2hg_{ab}$ in these coordinates and using (\ref{Connection-MetricCompatible}) we eventually find
\begin{equation*}\label{Explicit-CCCKV-Eq}
  2\,h\,g_{ab} \eq  (\partial_\lambda N) \, \delta^\lambda_{a}\delta^\lambda_{b} \ma \frac{1}{2}\left[\, \partial_\lambda g_{ab} \me \partial_a g_{b\lambda}
  \me \partial_b g_{a\lambda} \ma 2\,T_{(ab)\lambda} \, \right]\,.
\end{equation*}
Apart from the relations already obtained, the above equation is equivalent to
\begin{equation}\label{gij-constraint}
  \partial_\lambda \, g_{ij} \eq 4\,h\, g_{ij} \me 2 \,T_{(ij)\lambda} \,.
\end{equation}
Therefore, we have proved that a manifold endowed with a metric-compatible connection admits a vector field $\bl{H}$ obeying the CCCKV equation, $\nabla_aH_e=2hg_{ab}$, with $h\neq0$ if, and only if, its line element can be written as (\ref{Metric1}) with $N$ being a function of $\lambda$, $g_{ij}$ obeying the differential equation (\ref{gij-constraint}) and the torsion obeying the constraint (\ref{Tlambda ab}). In the previous coordinates the CCCKV is given by $\bl{H}=\bl{\partial}_\lambda$.

\subsection{The Case of a Totally Skew-Symmetric Torsion}

Let us now consider the important special case  of a totally skew-symmetric torsion, $T_{abc} = T_{[abc]}$. In such a case the Eq. (\ref{gij-constraint}) can be nicely integrated for the metric $g_{ij}$. Indeed, since in this case $T_{(ab)c} =0$, equations (\ref{gij-constraint}) and (\ref{PDN}) yield
$$ N\, \partial_\lambda \, g_{ij} \eq  g_{ij} \,\partial_\lambda N  \quad \Longrightarrow \quad  g_{ij}(\lambda,x) = N(\lambda)\, \tilde{g}_{ij}(x)\,, $$
where the functions $\tilde{g}_{ij}$ are functions just of the coordinates $\{x^i\}$. Therefore, the metric (\ref{Metric1}) can be written as
\begin{equation}\label{MetricSkew}
   ds^2 \eq N(\lambda)\,\left[\,d\lambda^2 \ma \tilde{g}_{ij}(x)\,dx^i dx^j \,\right] \,.
\end{equation}
Thus, a manifold endowed with a metric-compatible connection with totally skew-symmetric torsion admits a vector field $\bl{H}$ obeying the CCCKV equation if, and only if, its metric can be written as (\ref{MetricSkew}) and the torsion is such that $T_{\lambda ab}=0$. In the latter case the CCCKV is given by $\bl{H}=\bl{\partial}_\lambda$. It is interesting noting that, in the case of a totally skew-symmetric torsion, Eqs. (\ref{h'-prop-H}) and (\ref{h'-CCCKV}) imply that $\bl{H}$ is an eigenvector of the Ricci tensor, $R_a^{\ph{a}e}H_e\propto H_a$.

Now, suppose that a manifold admits a CCCKV $\bl{H}$ and a Killing vector $\bl{\eta}$. Then, let us prove that in this case the vector field $\chi_b=\nabla_b(H^c\eta_c)$ is a conformal Killing vector. Indeed, using (\ref{CCCKV}) we have that
$$ \chi_b\eq\nabla_b(H^c\eta_c) \eq 2 \,h\, \eta_b \ma H^c\, \nabla_b \eta_c \quad \Rightarrow\quad $$

\begin{equation}\label{Dchi1}
  \nabla_a  \chi_b \eq 2\,h'_a\,\eta_b \ma H^c \nabla_a \nabla_b \eta_c \,.
\end{equation}
Where it has been used the fact that $\bl{\eta}$ obeys the Killing equation.\footnote{Recall that in the case of a totally skew-symmetric torsion the two traditional definitions of Killing vector coincide, so that it is not necessary to specify what it is meant by a Killing vector (see section \ref{SubSec-KillingV}).} Now, inserting Eqs. (\ref{h'-CCCKV}) and (\ref{KillingV-SecondD}) into (\ref{Dchi1}) and using the condition $H^eT_{e ab}=0$ we find that:
\begin{equation*}\label{Dchi2}
  \nabla_a  \chi_b \eq \frac{-\,1}{n-1}\,R_a^{\ph{a}e}\,H_e\,\eta_b \ma H^c\,\eta^e \left[ \frac{3}{2}\,R_{[abc]e}\ma R_{bcea} \right] \me \frac{1}{2}\,T^e_{\ph{e}ab}\,H^c\,\nabla_e\eta_c \,.
\end{equation*}
Then, using (\ref{Riemann-Prop}) and (\ref{Riemann-Prop2}) to rewrite the terms $R_{[abc]e}$ and $R_{bcea}$ in the latter equation we have
\begin{equation*}\label{Dchi3}
  \nabla_a  \chi_b \eq \frac{-\,1}{n-1}\,R_a^{\ph{a}e}\,H_e\,\eta_b \ma H^c\,\eta^e \left[ -\,\frac{3}{2}\,\nabla_{[a}T_{bc]e}  \ma \lef R_{eabc} \ma \nabla_{[e}T_{a]bc} \me \nabla_{[b}T_{c]ea} \rig \right] \me \frac{1}{2}\,T^e_{\ph{e}ab}\,H^c\,\nabla_e\eta_c \,.
\end{equation*}
Now, making use of (\ref{DT-H}) and manipulating the derivative on the last term of the above relation, we end up with
\begin{equation*}\label{Dchi4}
  \nabla_a  \chi_b \eq \frac{-\,1}{n-1}\,R_a^{\ph{a}e}\,H_e\,\eta_b \ma \eta^e\, R_{eab}^{\ph{eab}c}\,H_{c} \me  3\,h\,\eta^e\,T_{eab} \me   \frac{1}{2}\,T^e_{\ph{e}ab}\, \left( \,\chi_e \me \eta_c\,\nabla_e H^c \, \right)   \,.
\end{equation*}
Finally, using the integrability condition (\ref{Int-CCCKV-Algeb}) to rewrite the term $\eta^e\, R_{eab}^{\ph{eab}c}\,H_{c}$ lead us to the following result:
\begin{equation}\label{Dchi5}
  \nabla_a  \chi_b \eq \frac{-\,1}{n-1}\,g_{ab}\,\left( \eta^e\,R_{ec}\,H^c \right) \me  \frac{1}{2}\,\chi_e\,T^e_{\ph{e}ab}   \,.
\end{equation}
In particular, taking the symmetric part of (\ref{Dchi5}) we see that the right hand side is proportional to the metric. Therefore, the following theorem can be stated.\\
\\
\textbf{Theorem -}\, \emph{Let $(M,\bl{g})$ be a manifold endowed with a metric-compatible connection whose torsion is totally skew-symmetric. Then, if $\bl{\eta}$ is a Killing vector field in $M$ and $\bl{H}$ obeys $\nabla_{a}H_b=2\,h\,g_{ab}$ with $h\neq0$ then the vector field $\chi_a=\nabla_a(H^c\eta_c)$ is a conformal Killing vector. Moreover, if $\chi^eT_{eab}=0$ then $\bl{\chi}$ is covariantly closed, $\nabla_{[a}\chi_{b]}=0$.}\\
\\
The torsion-free version of this theorem was proved in \cite{Bat-KYn-1}.

\subsection{Covariantly Constant Vector Fields, the Case $h=0$}

Previously, we have considered a vector field $\bl{H}$ obeying $\nabla_{a}H_b=2\,h\,g_{ab}$ with $h$ being non-vanishing. Now, let us consider the special case $h=0$, namely when the vector field $\bl{H}$ is covariantly constant,
\begin{equation}\label{DH=0}
 \nabla_aH_b \eq 0\,.
\end{equation}
The integrability condition is this case is
$$ R_{abc}^{\ph{abc}e}\,H_{e}\eq 0 \,.  $$
In the general case $h\neq0$ we have made use of (\ref{Hh'}) to prove that $\bl{H}$ is orthogonal to a family of hyper-surfaces. However, Eq. (\ref{Hh'}) is trivial for the case $h=0$, so that we cannot arrive at the same conclusion in such a case. Instead, using (\ref{APDH}) we have that
\begin{equation}\label{APDH2}
  \partial_{[a}H_{b]} \eq \frac{1}{2}\,H_e\,T^e_{\ph{e}ab}  \quad\; \textrm{ and } \;\quad  H_{[c}\,\partial_{a}H_{b]} \eq \frac{1}{2}\,H_e\,T^e_{\ph{e}[ab}H_{c]} \,.
\end{equation}
Note that for vanishing torsion Eq. (\ref{APDH2}) not only implies that $\bl{H}$ is hyper-surface-orthogonal but also guarantees that it $\bl{H}$ is closed. Nevertheless, for a general torsion neither conclusions need to hold.

In order to attain these conclusions more explicitly, let us introduce coordinates $\{x^i,\lambda\}$ with $i,j\in\{1,2,\cdots,n-1\}$ and such that $\bl{H}=\bl{\partial}_\lambda$. In these coordinates the metric is generally written as
\begin{equation}\label{Metric3}
   ds^2 \eq N\,d\lambda^2 \ma 2\,\xi_i\,dx^i\,d\lambda \ma g_{ij}\,dx^i dx^j \,.
\end{equation}
Where $N$ is a constant while $\xi_i$ and $g_{ij}$ can, in principle, be general functions of $\lambda$ and $\{x^i\}$. Now, imposing the equation $\nabla_aH^b=0$ in this coordinate frame we find that the connection symbol $\Gamma^b_{a\lambda}$ must vanish, which is tantamount to
\begin{equation}\label{DiffEq-g-2}
 \partial_a g_{\lambda c} \ma  \partial_\lambda g_{ac} \me \partial_c g_{a\lambda} \eq T_{a\lambda c} \ma T_{\lambda a c} \me T_{ca\lambda } \,.
\end{equation}
This equality, in turn, is equivalent to the following three constraints:
\begin{equation}\label{Constraints-h0}
    \partial_\lambda \xi_i \eq T_{\lambda\lambda i}     \quad,\quad  \partial_\lambda g_{ij} \eq -\,2\,T_{(ij)\lambda} \quad,\quad \partial_i \xi_j \me  \partial_j \xi_i  \eq  T_{\lambda ij}\,.
\end{equation}
In particular, the third condition in (\ref{Constraints-h0}) guarantees that if $T_{\lambda ij}$ is non-vanishing then the functions $\xi_i$ cannot all vanish, so that generally $\bl{H}$ is not orthogonal to a family of hyper-surfaces. This is intriguing. Since the equation $\nabla_aH_b=0$ is more restrictive than the CCCKV equation,  $\nabla_{a}H_b=2\,h\,g_{ab}$, it is natural to expect that the results valid for $h\neq 0$  would also be valid in the more special case $h=0$. However, we have proved that there are constraints valid in the case $h\neq 0$ that do not carry to the special case $h=0$, such as the hyper-surface-orthogonal condition and the restriction $T_{\lambda ab}=0$.  Note, however, that such unexpected fact happens only when the torsion is non-zero.


\section{Maximally Symmetric Spaces}\label{Sec.MaximallySym}

A Riemannian manifold endowed with the Levi-Civita connection is called maximally symmetric when it admits the maximum number independent Killing vector fields. In $n$ dimensions this maximal number is $\frac{1}{2}n(n+1)$, which physically arises from $n$ translations an $\frac{1}{2}n(n-1)$ rotations. In this section, we shall consider the issue of defining a maximally symmetric space in the presence of torsion. Here, we shall say that $\bl{\eta}$ is a Killing vector field if it obeys the Killing equation
$$ \nabla_a\,\eta_b \ma  \nabla_b\,\eta_a \eq 0 \,. $$
Recall that we could have defined a Killing vector as a vector field such that $\mathcal{L}_{\bl{\eta}}g_{ab}=0$. As pointed out in Sec. \ref{SubSec-KillingV}, these two definitions are not equivalent in the presence of general torsion. However, since
$$  \mathcal{L}_{\bl{\eta}}g_{ab} =  \mathring{\nabla}_a\,\eta_b \ma  \mathring{\nabla}_b\,\eta_a \,,$$
it turns out that the latter definition is just a particular case of the former. Namely, the equation $\mathcal{L}_{\bl{\eta}}g_{ab}=0$ can be retrieved from the equation $\nabla_{(a}\,\eta_{b)}=0$ by choosing the connection to be torsion-free. Maximally symmetric spaces in the presence of torsion have been investigated before. In \cite{Gangop.}, a couple of integrability conditions for the Killing equation are worked out, but along the calculations some unjustified constraints are imposed over the torsion as well as over the curvature. In Refs. \cite{Multamaki:2008tk,Sur:2013aia}, it is imposed that the torsion tensor should also be ``maximally symmetric'', namely it is assumed that the torsion is invariant under the Lie dragging along the Killing vectors of a maximally symmetric manifold or submanifold, see also \cite{Tsamparlis}. Here we go further and deduce the integrability conditions of a maximally symmetric space without imposing any constraint over the torsion. Moreover, we also investigate spaces with the maximal number of KY tensors of order $n-1$. For a nice review of maximally symmetric spaces in the torsion-free case, the reader is referred to Ref. \cite{Weinberg-Book}.

The first natural question is: what is the maximum number of Killing vectors when the connection has non-zero torsion? The answer is held in Eq. (\ref{KillingV-SecondD}), according to which the second derivative of a Killing vector can be written in terms of the vector itself and its first derivative. Thus, successively differentiating this equation one can write all derivatives of the Killing vector in terms of the Killing vector itself and its first derivative. So, if the components of $\eta_a$ and $\nabla_a\eta_b$ are given in one point of the manifold it is possible to reconstruct the vector field $\bl{\eta}$ in the whole manifold. Therefore, since $\eta_a$ has $n$ components and $\nabla_a\eta_b$ is an antisymmetric rank 2 tensor with $\frac{1}{2}n(n-1)$ components, it follows that maximum number of Killing vector fields remains being $\frac{1}{2}n(n+1)$. Therefore, even in the presence of torsion, we shall say that an $n$-dimensional manifold is maximally symmetric if it admits $\frac{1}{2}n(n+1)$ independent Killing vector fields.

Now, let us investigate how the existence of $\frac{1}{2}n(n+1)$ independent Killing vector fields constrains the metric and the torsion of a manifold. In order to accomplish this, we should find the integrability condition for the Killing equation. Using the identity (\ref{Riemann-Def}) we have that
$$  \lef \nabla_d \nabla_a  \me \nabla_a \nabla_d \rig \nabla_b \eta_c \eq R_{dab}^{\ph{dab}e} \, \nabla_e\eta_c \ma  R_{dac}^{\ph{dac}e} \, \nabla_b\eta_e \me T^e_{\ph{e}da}\,\nabla_e\nabla_b\eta_c\ $$
Now, rewriting the second derivative of $\bl{\eta}$ in the right hand side of the above equation by means of (\ref{KillingV-SecondD}) we obtain
\begin{align}
  2 \lef \nabla_d\nabla_a\nabla_b\eta_c -\nabla_a\nabla_d\nabla_b \eta_c \rig   \eq&  2 R_{dabe} \nabla^{e}\eta_{c} \ma  2\, R_{dace} \nabla_{b}\eta^{e} \ma R_{bcef}\, T^{f}{}_{ad} \eta^{e} \me  R_{bfce} \, T^{f}{}_{ad} \eta^{e}   \nonumber\\ \ma R_{cfbe} \, T^{f}{}_{ad} \eta^{e}
& \ma T_{ebf} \, T^{f}{}_{ad} \nabla^{e}\eta_{c} \ma
T_{ead} \, T_{fbc} \nabla^{f}\eta^{e}  \me T_{ecf}\, T^{f}{}_{ad} \nabla^{e}\eta_{b} \label{DDDet1}
\end{align}
Then, using  (\ref{KillingV-SecondD}) to write the terms $\nabla_a\nabla_b\eta_c$ and $\nabla_d\nabla_b\eta_c$ on the left hand side of (\ref{DDDet1}) and then using (\ref{KillingV-SecondD}) again to rewrite the second derivatives of $\bl{\eta}$ that eventually appear, lead us to the following integrability condition:
\begin{gather}
  \left(\, R_{cdef} \,T^{f}{}_{ab} - R_{cfde} \,T^{f}{}_{ab}  -
R_{dfce}\, T^{f}{}_{ab}  - R_{bdef} \,T^{f}{}_{ac}  +
R_{bfde}\, T^{f}{}_{ac}  +   R_{dfbe} \,T^{f}{}_{ac}   - 2
R_{bcef}\, T^{f}{}_{ad} \right. \nonumber\\
+  2 R_{bfce}\, T^{f}{}_{ad}  - 2 R_{cfbe}\, T^{f}{}_{ad}  - 2 \,R_{adef} \,T^{f}{}_{bc}  -
R_{acef}\, T^{f}{}_{bd} - R_{afce} \,T^{f}{}_{bd}   - R_{cfae}\, T^{f}{}_{bd}  +   R_{abef}\, T^{f}{}_{cd}  \nonumber\\
+ R_{afbe}\, T^{f}{}_{cd} +   \left. R_{bfae}\, T^{f}{}_{cd}  +
2 \nabla_{a}R_{bcde} +  2 \nabla_{a}R_{bdce} - 2
 \nabla_{a}R_{cdbe} +  2 \nabla_{d}R_{abce} - 2
 \nabla_{d}R_{acbe} -2 \nabla_{d}R_{bcae}\, \right) \eta^{e} \nonumber\\
  \,=\,   - 2\, R_{bcde} \nabla_{a}\eta^{e} - 2\, R_{bdce} \nabla_{a}\eta^{e} + 2\, R_{cdbe} \nabla_{a}\eta^{e} - 4\,R_{adce} \nabla_{b}\eta^{e}
   - 2\, R_{abce} \nabla_{d}\eta^{e} + 2 R_{acbe} \nabla_{d}\eta^{e}  \label{Int.KillingV-1} \\
   + 2 R_{bcae} \nabla_{d}\eta^{e} -  T_{ecf} T^{f}{}_{bd} \nabla^{e}\eta_{a} + T_{ebf} T^{f}{}_{cd} \nabla^{e}\eta_{a} - 2 \nabla_{d}T_{ebc} \nabla^{e}\eta_{a} + T_{edf} T^{f}{}_{ac} \nabla^{e}\eta_{b} - 2\, T_{ecf} T^{f}{}_{ad} \nabla^{e}\eta_{b} \nonumber\\ +
T_{eaf} T^{f}{}_{cd} \nabla^{e}\eta_{b}  - 2 \nabla_{a}T_{ecd} \nabla^{e}\eta_{b} - 2 \nabla_{d}T_{eac} \nabla^{e}\eta_{b} - 4
R_{adbe} \nabla^{e}\eta_{c} -  T_{edf} T^{f}{}_{ab} \nabla^{e}\eta_{c} + 2 T_{ebf} T^{f}{}_{ad} \nabla^{e}\eta_{c} \nonumber\\ -
T_{eaf} T^{f}{}_{bd} \nabla^{e}\eta_{c} + 2 \nabla_{a}T_{ebd} \nabla^{e}\eta_{c}  + 2 \nabla_{d}T_{eab} \nabla^{e}\eta_{c} -  T_{ecf}
T^{f}{}_{ab} \nabla^{e}\eta_{d} + T_{ebf} T^{f}{}_{ac} \nabla^{e}\eta_{d} + 2 \nabla_{a}T_{ebc} \nabla^{e}\eta_{d} \nonumber\\-  T_{ecd}
T_{fab} \nabla^{f}\eta^{e} + T_{ebd} T_{fac} \nabla^{f}\eta^{e}  + 2
T_{ebc} T_{fad} \nabla^{f}\eta^{e} + 2\, T_{ead} T_{fbc} \nabla^{f}\eta^{e} + T_{eac} T_{fbd} \nabla^{f}\eta^{e} -  T_{eab}
T_{fcd} \nabla^{f}\eta^{e} \,.\nonumber
\end{gather}
This integrability condition must hold for any vector field obeying the Killing equation. In the particular case of a maximally symmetric space, we can make linear combinations of the $\frac{1}{2}n(n+1)$  Killing vectors in order to construct a Killing vector field such that $\eta_a$ has arbitrary components at some  point of the manifold and such that $\nabla_a\eta_b$ is an arbitrary skew-symmetric \textit{matrix} at the same point. In particular, we can take $\nabla_a\eta_b=0$ and $\eta_a=\delta_a^e$, with $e$ being an arbitrary fixed coordinate index. In this case we are led to the following integrability condition:
\begin{gather}
 R_{cdef} \,T^{f}{}_{ab} - R_{cfde} \,T^{f}{}_{ab}  -
R_{dfce}\, T^{f}{}_{ab}  - R_{bdef} \,T^{f}{}_{ac}  +
R_{bfde}\, T^{f}{}_{ac}  +   R_{dfbe} \,T^{f}{}_{ac}   - 2
R_{bcef}\, T^{f}{}_{ad}  \nonumber\\
+  2 R_{bfce}\, T^{f}{}_{ad}  - 2 R_{cfbe}\, T^{f}{}_{ad}  - 2 \,R_{adef} \,T^{f}{}_{bc}  -
R_{acef}\, T^{f}{}_{bd} - R_{afce} \,T^{f}{}_{bd}   - R_{cfae}\, T^{f}{}_{bd}  +   R_{abef}\, T^{f}{}_{cd}  \label{Int.KillingV-et}\\
+ R_{afbe}\, T^{f}{}_{cd} +    R_{bfae}\, T^{f}{}_{cd}  +
2 \nabla_{a}R_{bcde} +  2 \nabla_{a}R_{bdce} - 2
 \nabla_{a}R_{cdbe} +  2 \nabla_{d}R_{abce} - 2
 \nabla_{d}R_{acbe} -2 \nabla_{d}R_{bcae} \eq 0  \,. \nonumber
\end{gather}
Differently, we can take $\eta_a=0$ and $\nabla_a\eta_b=2\,\delta_{[a}^h\delta_{b]}^k$ in some arbitrary point of the manifold with $h$ and $k$ being arbitrary fixed indices. In the latter case, Eq. (\ref{Int.KillingV-1}) lead us to the following constraint that must be satisfied by a maximally symmetric space:
\begin{gather}
 4 g_{ck} R_{dabh} + 4 g_{ch} R_{adbk} + 4 g_{bk} R_{adch} - 4 g_{bh}
R_{adck} + 2 g_{ak} R_{bcdh} - 2 g_{ah} R_{bcdk} + 2 g_{ak} R_{bdch}
- 2 g_{ah} R_{bdck} \nonumber \\ - 2 g_{ak} R_{cdbh} + 2 g_{ah} R_{cdbk} -  g_{ck}
T_{fbd} T_{ha}{}^{f} + g_{bk} T_{fcd} T_{ha}{}^{f} + 2 g_{ck} T_{fad}
T_{hb}{}^{f} + g_{ak} T_{fcd} T_{hb}{}^{f} - 2 g_{bk} T_{fad} T_{hc}{}^{f} \nonumber\\
 -  g_{ak} T_{fbd} T_{hc}{}^{f} -  g_{ck} T_{fab}
T_{hd}{}^{f} + g_{bk} T_{fac} T_{hd}{}^{f} + g_{ch} T_{fbd}
T_{ka}{}^{f} -  g_{bh} T_{fcd} T_{ka}{}^{f} - 2 g_{ch} T_{fad} T_{kb}{}^{f} -  g_{ah} T_{fcd} T_{kb}{}^{f}  \label{Int.KillingV-Det} \\    + 2 g_{bh} T_{fad}T_{kc}{}^{f}
+ g_{ah} T_{fbd} T_{kc}{}^{f} + g_{ch} T_{fab} T_{kd}{}^{f} -  g_{bh} T_{fac} T_{kd}{}^{f}   + 2 g_{ck} \nabla_{a}T_{hbd} - 2 g_{bk} \nabla_{a}T_{hcd}   -
2 g_{ch} \nabla_{a}T_{kbd} \nonumber\\+ 2 g_{bh} \nabla_{a}T_{kcd} + 2 g_{ck} \nabla_{d}T_{hab}- 2 g_{bk} \nabla_{d}T_{hac} - 2 g_{ak} \nabla_{d}T_{hbc} - 2 g_{ch} \nabla_{d}T_{kab} + 2 g_{bh} \nabla_{d}T_{kac} + 2 g_{ah} \nabla_{d}T_{kbc}   \nonumber\\
 + g_{dk} (2 R_{abch} - 2 R_{acbh} - 2 R_{bcah} + T_{fac} T_{hb}{}^{f}  -  T_{fab} T_{hc}{}^{f}+ 2 \nabla_{a}T_{hbc})  \;\quad \nonumber\\
 + g_{dh} (-2 R_{abck} + 2 R_{acbk}  + 2 R_{bcak} - T_{fac} T_{kb}{}^{f} + T_{fab} T_{kc}{}^{f}   - 2 \nabla_{a}T_{kbc}) \eq 0 \,. \nonumber
\end{gather}
In particular, contracting the latter equation with $g^{bk}$ we find the following simpler condition which can be proved to be equivalent to (\ref{Int.KillingV-Det}):
\begin{gather}
  (n-1) \lef 4\, R_{adch} \ma T_{fcd}\, T_{ha}{}^{f}   \me T_{fca} T_{hd}{}^{f} \me 2  \,T_{fad} \,T_{hc}{}^{f}  \ma 2\,  \nabla_{a}T_{hdc} \me  2  \nabla_{d}T_{hac}\rig \nonumber\\
    \me g_{dh} (2\, R_{ac} \ma 2\, R_{ca}  \me  T^{f}{}_{a}{}^{k} T_{kcf} \me  T^{f}{}_{ac} T^{k}{}_{fk} \me  2 \nabla_{a}T^{f}{}_{cf}) \nonumber\\
  \ma  g_{ah} (2\, R_{cd} \ma  2\, R_{dc} \me  T^{f}{}_{c}{}^{k} T_{kdf} \ma T^{f}{}_{cd} T^{k}{}_{fk} \me 2 \nabla_{d}T^{f}{}_{cf}) \label{Int.KillingV-Det2}\\
  + 2\, g_{ch} \lef  T^{f}{}_{ad}\, T^{k}{}_{fk} +  \nabla_{a}T^{f}{}_{df} -  \nabla_{d}T^{f}{}_{af} \rig \eq 0 \,. \nonumber
 \end{gather}
Therefore, the integrability conditions satisfied by a maximally symmetric space are (\ref{Int.KillingV-et}) and (\ref{Int.KillingV-Det2}). Now, contracting (\ref{Int.KillingV-Det2}) with $g^{ac}$ and using the curvature properties (\ref{Riemann-Prop}) we conclude that in a maximally symmetric space the Ricci tensor might be given by:
\begin{align}
 4 n \,R_{dh} \eq& g_{dh} \lef 4\, R -  T^{abf} T_{baf} + 2 \,\nabla_{b}T^{a}{}_{a}{}^{b} \rig
 + (n-1) \lef T_{adb} T_{h}{}^{ab}  - 2\,\nabla_{a}T_{hd}{}^{a} \rig + 2\, \nabla_{a}T^{a}{}_{dh}   \nonumber\\
 & \ma T^{a}{}_{d}{}^{b} T_{bha} + 5\, T^{a}{}_{dh} T^{b}{}_{ab}   + 6\, \nabla_{d}T^{a}{}_{ha} - 4 \,\nabla_{h}T^{a}{}_{da} \,  \label{Int.KillingV-Ricci}
\end{align}
with $R\equiv g^{ab}R_{ab}$ denoting the Ricci scalar. Then, inserting the above expression for the Ricci tensor into the integrability condition (\ref{Int.KillingV-Det2}) lead us to the following expression for the curvature tensor:
\begin{gather}
R_{adch} \eq  \frac{1}{8n(n-1)} \left\{ (g_{ac} g_{dh} -  g_{ah} g_{cd} ) (8 R - 2\, T^{bfk} T_{fbk} + 4 \nabla_{f}T^{b}{}_{b}{}^{f})
+ 4 n g_{ch} ( 2\,\nabla_{[d}T^{b}{}_{a]b}  -  T^{b}{}_{ad} T^{f}{}_{bf} )  \right. \nonumber\\
 + g_{dh} \left[  (n-1) \left( T_{a}{}^{bf} T_{bcf} \ma  T_{baf} T_{c}{}^{bf}  - 2 \, T^{b}{}_{a}{}^{f} T_{fcb} - 4  \nabla_{b}T_{(ca)}{}^{b} \right) \right. \nonumber \\
\left. \quad\; + 2 n  T^{b}{}_{ca} T^{f}{}_{bf}   + 2  \nabla_{c}T^{b}{}_{ab}  + 2 \nabla_{a}T^{b}{}_{cb} - 4 n \nabla_{a}T^{b}{}_{cb} \,\right] \nonumber\\
  - g_{ah} \left[ (n-1) \left( T_{bdf} T_{c}{}^{bf}  +  T_{bcf} T_{d}{}^{bf}   - 2  T^{b}{}_{c}{}^{f} T_{fdb} - 4  \nabla_{b}T_{(cd)}{}^{b} \right) \right.
  \label{Int.KillingV-Riemann}\\
  \left. \quad\; + 2 n \ T^{b}{}_{cd} T^{f}{}_{bf}    + 2 \nabla_{c}T^{b}{}_{db} + 2 \nabla_{d}T^{b}{}_{cb} - 4 n \nabla_{d}T^{b}{}_{cb} \, \right] \nonumber\\
\left. + 2\, n (n-1)  \left( 2 \nabla_{a}T_{hcd} + 2 \nabla_{d}T_{hac} + 2 \,T_{bad} T_{hc}{}^{b} - T_{bcd} T_{ha}{}^{b}  -  T_{bac} T_{hd}{}^{b} \right) \right\}  \nonumber \,.
\end{gather}
Note that while the left hand side of (\ref{Int.KillingV-Riemann}) is skew-symmetric in the indices $ch$, the right hand side is not automatically skew-symmetric. Therefore, taking the symmetric part of the equation (\ref{Int.KillingV-Riemann}) in the indices $ch$ we are led to a first order differential equation for the torsion. Particularly, contracting (\ref{Int.KillingV-Riemann}) with $g^{ch}$ we find that
$$ \nabla_a\,T^e_{\ph{e}de} \me  \nabla_d\,T^e_{\ph{e}ae}  \eq  \me T^e_{\ph{e}ke}\,T^k_{\ph{k}ad}  \,.$$
It is worth mentioning that the latter constraint is necessary in order to prove the equivalence between (\ref{Int.KillingV-Det}) and (\ref{Int.KillingV-Det2}).

Using (\ref{Riemann-LeviC}) to  write the curvature tensor on the left hand side of (\ref{Int.KillingV-Riemann}) in terms of the Levi-Civita Riemann tensor lead us to a second order differential equation for the metric with the torsion being the source. Then, once we have found the metric we can substitute the expression (\ref{Int.KillingV-Riemann}) for the curvature tensor into the integrability condition (\ref{Int.KillingV-et}), which yields a second order differential equation for the torsion. Going through these steps in the case of general torsion is certainly rather involved. So, probably, the best way to address these constraints is to make some simplifying assumptions on the torsion. Particularly, in the special case of a totally skew-symmetric torsion, it turns out that all these integrability conditions can be solved. Indeed, in this case using the integrability condition (\ref{Int.KillingV-Riemann}) along with (\ref{Riemann-LeviC}) lead us to the following equation:
\begin{equation}\label{Riemann-KVAT}
  T_{abc}\eq T_{[abc]}  \quad \quad \Longrightarrow \quad \quad \mathring{R}_{adch} \eq \frac{\mathring{R}}{n\,(n-1)}\,\lef g_{ac}g_{dh}\me g_{ah}g_{dc} \rig \,,
\end{equation}
with $\mathring{R}\equiv g^{ac} g^{bd} \mathring{R}_{abcd}$ denoting the Ricci scalar of the Levi-Civita curvature. In particular, the second Bianchi identity implies that $\mathring{R}$ is constant. Moreover, if (\ref{Riemann-KVAT}) holds it turns out that the integrability condition (\ref{Int.KillingV-et}) is automatically satisfied. Therefore, we conclude that the integrability conditions of a maximally symmetric space, namely Eqs. (\ref{Int.KillingV-et}) and (\ref{Int.KillingV-Det2}), impose no constraint over the torsion tensor if the torsion is  totally skew-symmetric. The spaces obeying Eq. (\ref{Riemann-KVAT}) are the well-known maximally symmetric spaces according to the Levi-Civita connection. For instance, in the case of Euclidean signature the metrics compatible with (\ref{Riemann-KVAT}) can always be written in suitable coordinates $\{x^a\}$ as
\begin{equation}\label{Metric-MaxSym}
  g_{ab} \eq \frac{1}{(1\ma \kappa\,r^2)^2}\,\delta_{ab}\;, \; \textrm{ where } \quad  r^2\eq \left[\, (x^1)^2\ma(x^2)^2\ma\cdots\ma(x^n)^2  \,\right]\,,\;\;\textrm{ and }\;\;  \kappa\,=\, \textrm{constant}\,.
\end{equation}
One can grasp this interesting result regarding maximally symmetric spaces with respect to connections with totally skew-symmetric torsion by analysing Eq. (\ref{Killings-Torsion}), according to which
$$  T_{abc}\eq T_{[abc]}  \quad \quad \Longrightarrow \quad \quad \nabla_a\,\eta_b \ma  \nabla_b\,\eta_a  \eq  \mathring{\nabla}_a\,\eta_b \ma  \mathring{\nabla}_b\,\eta_a \,. $$
Therefore, a vector field $\bl{\eta}$ obeys the Killing equation for a connection with totally anti-symmetric torsion if, and only if, it obeys the Killing equation for the Levi-Civita connection. Thus, if a manifold $(M,\bl{g})$ is maximally symmetric according to the Levi-Civita connection then this manifold is also maximally symmetric with respect to any connection whose torsion is totally skew-symmetric. Moreover, the Killing vectors are the same for both connections. For instance, using the coordinates adopted in (\ref{Metric-MaxSym}), the $\frac{1}{2}n(n+1)$ independent Killing vectors in the Euclidean case are given by
\begin{equation*}\label{Killing-MS}
\bl{\eta}^{i} \eq (1-\kappa\,r^2)\bl{\partial}_i \ma 2\,\kappa\, x^i\,x^a\bl{\partial}_a    \quad    \textrm{ and } \quad \bl{\eta}^{ij} \eq x^i\,\bl{\partial}_j \me x^j\,\bl{\partial}_i   \,,
\end{equation*}
where $i,j$ are labels running from 1 to $n$ and $i<j$.

\subsection{Maximal Number of Killing-Yano Tensors of Order $n-1$}

One remarkable property of the maximally symmetric manifolds with respect to the Levi-Civita connection is that, besides admitting the maximal number of Killing vectors, they also admit the maximal number of hidden symmetries, namely Killing tensors and Killing-Yano tensors. So, it is natural to wonder whether this property remains valid for maximally symmetric manifolds with respect to connections with non-zero torsion. In this section, we shall see that the answer is negative. More precisely, by means of analysing the simpler case of a Killing-Yano tensor of order $n-1$, we shall prove that in general a maximally symmetric space will not admit the maximal number of Killing-Yano tensors when the torsion is different from zero.

As mentioned before, every Killing-Yano tensor of rank $p$ is the Hodge dual of a covariantly closed conformal Killing-Yano tensor of rank $(n-p)$ and \textit{vice-versa}. Therefore, instead of considering Killing-Yano tensors of order $n-1$, we shall deal with covariantly closed conformal Killing vectors, \textit{i.e.}, vector fields $\bl{H}$ obeying (\ref{CCCKV}). Taking the covariant derivative of (\ref{CCCKV}) and then using (\ref{h'Algeb}) lead us to:
\begin{equation}\label{DDH}
 \nabla_a\nabla_b\,H_c \eq 2\,g_{bc}\,\nabla_a h \eq  \frac{-\,1}{n-1}\,g_{bc}\, \left( \, R_a^{\ph{a}e}\,H_e \me 2\,h\,T^e_{\ph{e}ea} \right) \,.
\end{equation}
Analogously, differentiating the latter equation and using (\ref{h'Algeb}) one find that all derivatives of $\bl{H}$ can be written in terms of $\bl{H}$ and  $h=\frac{1}{2n}\nabla^a H_a$. Thus, if we know the values of $H_a$ and $\nabla^aH_a$ at some point of the manifold then we can obtain $H_a$ throughout the whole manifold. Therefore, the maximal number of independent covariantly closed conformal Killing vectors in an $n$-dimensional manifold is $(n+1)$. In addition, if a manifold admits the maximal number of CCCKVs then it is always possible to find a CCCKV $\bl{H}$ such that $H_a$ and $\nabla^aH_a$ have any desired value at some arbitrary point of the manifold. Hence, integrability condition (\ref{Int-CCCKV-Algeb}) implies that if a manifold admits the maximal number of covariantly closed conformal Killing vectors then the following constraints hold:
\begin{equation}\label{MaxCCCKV}
   \left\{
  \begin{array}{cl}
   (n-1)R_{abc}^{\ph{abc}e} \me g_{ac}\,R_b^{\ph{b}e} \ma g_{bc}\,R_a^{\ph{a}e} \eq 0  \\
    (n-1)\,T_{cab}\me g_{ac}\,T^e_{\ph{e}eb} \ma  g_{bc}\,T^e_{\ph{e}ea}  \eq 0  \,. \\
    \end{array}
\right.
\end{equation}
Particularly, the second of these constraints implies that the torsion has the following form
\begin{equation}\label{MaxCCCKV-Torsion}
   T_{cab} \eq \frac{2}{n-1}\,T^e_{\ph{e}e[b}g_{a]c} \,.
\end{equation}
Note, in particular, that the totally skew-symmetric part of the torsion must be identically zero. This fact contrasts with what we have seen about spaces admitting the maximum number of Killing vectors. While the existence of the maximum number of CCCKVs implies that $T_{[abc]}$ must vanish, the existence of the maximum number of Killing vectors imposes no constraint over the torsion when it is totally antisymmetric.

Now, working out the first integrability condition in (\ref{MaxCCCKV}) lead us to the following expression for the curvature tensor:
\begin{equation}\label{MaxRiemann1}
 R_{abcd} \eq \frac{R}{n(n-1)}\, \lef g_{ac}\,g_{bd} \me g_{ad}\, g_{bc}  \rig \,.
\end{equation}
Then, using (\ref{Riemann-LeviC}) to rewrite the curvature in the above expression in terms of the Levi-Civita curvature and then using (\ref{MaxCCCKV-Torsion}), eventually lead us to the following expression for the Levi-Civita curvature:
\begin{equation}\label{MaxRiemann2}
 \mathring{R}_{abcd} \eq \frac{\mathring{R} \ma 2\,\nabla_e T^{ke}_{\ph{ke}k}}{n(n-1)}\, \lef g_{ac}\,g_{bd} \me g_{ad}\, g_{bc}  \rig \ma
\frac{2}{n-1}\lef  g_{d[a}\nabla_{b]} T^{e}_{\ph{e}ce} \me   g_{c[a}\nabla_{b]} T^{e}_{\ph{e}de}  \rig  \,.
\end{equation}
The above condition provides a second order differential equation for the metric, with the torsion being the source. Then, once we have found the metric, Eq. (\ref{MaxCCCKV-Torsion}) yields a first order differential equation for the torsion. In addition, note that atisymmetrizing the indices $abc$ in (\ref{MaxRiemann2}) and using the first Bianchi identity satisfied by the Levi-Civita Riemann tensor yields the constraint $g_{d[a}\nabla_{b} T^{e}_{\ph{e}c]e}=0$, which is equivalent to $\nabla_{[a}T^{e}_{\ph{e}b]e}=0$ if $n>2$.


\section{Conclusions}

In the present article the integrability conditions for the existence of Killing-Yano tensors, of arbitrary order, with respect to general metric-compatible connections have been obtained. Thus, extending the results of Refs. \cite{Kashiwada-Int,Bat-IntegraB-CKY} to the case of non-zero torsion. As we have seen, the case of totally skew-symmetric torsion is of special appeal, since in such a case the concepts of geodesic and Killing vector are not ambiguous. Moreover, in superstring theory the torsion is associated to the field strength of the Kalb-Ramond field, so that it is automatically antisymmetric. In spite of such motivations, for sake of completeness, here no restriction over the torsion has been assumed. Once the curvature of the connection is known, the algebraic integrability conditions found here can be used to eliminate several components of the Killing-Yano tensor, facilitating the integration of the Killing-Yano equation. Indeed, this kind of procedure have been successfully applied in Ref. \cite{Houri:2014} for the torsion-free case. As an application of the results obtained here, in Sec. \ref{Sec.CCCKV} all metrics and torsions compatible with the existence of a Killing-Yano tensor of order $n-1$ have been found.

It is worth pointing out the important role played by KY tensors in quantum field theory in curved spacetimes. Although symmetric Killing tensors generate conserved charges at the classical level, it turns out that at the quantum level an anomaly term involving the curvature pops up. So, generally, Killing tensors do not yield a conservation law in the quantum theory \cite{Santillan}. Differently, in the absence of torsion, KY tensors yield conserved charges for the Klein-Gordon and Dirac equations both at classical and quantum levels \cite{Cariglia,Santillan}. Nevertheless, in Ref. \cite{Houri:TorsionDirac} it has been proved that a skew-symmetric torsion generally spoils this property of KY tensors, namely in the quantum theory an anomaly involving the torsion arises. Actually, an anomaly due to the torsion shows up even at the semi-classical treatment of a spinning particle \cite{Houri:TorsionDirac}.

Here, the issue of defining a maximally symmetric space in the presence of torsion has also been addressed. Differently from the previous works on this topic, in the present article we have not assumed that the torsion is invariant by the isometries of the space, making the approach adopted here more general. It has been obtained the restrictions that the curvature and the torsion must obey in order for the space to admit the maximum number of vector fields obeying the Killing equation with respect to a general metric-compatible connection. Particularly, it has been proved that in the case of a totally skew-symmetric torsion the metric of a maximally symmetric space must be the same metric of the torsion-free case and that no restriction is imposed over the torsion. Moreover, we have shown that, contrary to the torsion-less case, a maximally symmetric space in the presence of torsion generally does not admit the maximum number of Killing-Yano tensors. Hopefully, these results regarding maximally symmetric spaces can be valuable for the study of cosmological models in the presence of torsion.

\newpage

\section*{Acknowledgments}
I really want to thank Tsuyoshi Houri for suggesting the theme of this article as well as for valuable discussions. I also thank the Brazilian funding agency CAPES (Coordena\c{c}\~{a}o de Aperfei\c{c}oamento de Pessoal de N\'{\i}vel Superior) for part of the financial support.


\end{document}